\pdfoutput=1
\documentclass[journal]{IEEEtran}
\usepackage{amssymb}
\usepackage{graphicx}
\usepackage{graphics}
\usepackage{array} 
\usepackage{ctable}
\usepackage{multirow}
\usepackage[footnotesize]{subfigure}
\usepackage{upgreek}
\usepackage{cite}
\usepackage{color}
\usepackage{amsmath}
\usepackage{amsfonts}
\usepackage{subfigure}
\usepackage{threeparttable} 
\usepackage{booktabs}
\usepackage{makecell}
\ifCLASSINFOpdf
\else
\fi
\begin{document}
%
\title{Improved Source Counting and Separation\\ for Monaural Mixture}

%
\author{Yiming~Xiao 
        and~Haijian~Zhang
        \\Signal Processing Lab., School of Electronic Information, Wuhan University, China
       \vspace{-1.3\baselineskip}
}
\markboth{Time-frequency analysis and its applications}
{}
\maketitle

\begin{abstract}
Single-channel speech separation in time domain and frequency domain has been widely studied for  voice-driven applications over the past few years. Most of previous works assume known number of speakers in advance, however,  which is not easily accessible through monaural mixture in practice. In this paper, we propose a novel model of single-channel multi-speaker separation by jointly learning the  time-frequency feature and the unknown number of speakers. Specifically, our model integrates the time-domain convolution encoded feature map and the frequency-domain spectrogram by attention mechanism, and the integrated features are projected into high-dimensional embedding vectors which are then  clustered with deep attractor network to modify the  encoded feature. Meanwhile, the number of speakers is counted by computing the Gerschgorin disks of the embedding vectors which are orthogonal for different speakers. Finally, the modified encoded feature is inverted to the sound waveform using a linear decoder.  Experimental evaluation on the GRID dataset shows that  the proposed method with a single model can accurately estimate the number of speakers with 96.7 \% probability of success, while achieving the state-of-the-art separation results on multi-speaker mixtures in terms of scale-invariant signal-to-noise ratio improvement (SI-SNRi) and signal-to-distortion ratio improvement (SDRi).  
\end{abstract}


\section{Introduction}
\IEEEPARstart{M}{onaural} speech separation with pretty performance is an important prerequisite for robust speech processing in real-world acoustic environments. For instance, automatic speech recognition (ASR) in multi-speaker conditions first requires the separation of individual speakers from their monaural mixture before identifying a target speaker or recognizing target speech \cite{QIAN20181,KHADEMIAN20181}. The well-known cocktail party problem  which is effortless for humans has been shown to be difficult for computer algorithms \cite{8369155,8683850,7979557,Nie2015TwostageMJ,fan2020deep}. Therefore, substantial efforts should be made to solve the cocktail-party problem based solely on a monaural mixture. 

To tackle this problem, various methods including computational auditory scene analysis (CASA) \cite{1709891,403c9e27c60d4acaa65c7720b40b949e,7886000}, non-negative matrix factorization (NMF) \cite{6854302,6854305,5495567} were proposed. Recently, the major development in deep learning techniques has led to a big step forward in solving the speech separation task. Most deep learning based techniques were studied in frequency domain \cite{8462471,Wang2019,Liu2019,inproceedings,7471631,Isik+2016,7952155,8264702,8462507}, in which the deep clustering (DC) approach  projected the mixture spectrogram to a high-dimensional embedding space which was more discriminative for speaker partitioning \cite{7471631,Isik+2016}. Based on the DC,  deep attractor network (DANet) \cite{7952155,8264702} or loss functions \cite{8462507}  were introduced to improve the separation performance. In addition, Luo \textit{et al.}  \cite{8462116,8707065} introduced a new solution to speech separation in time domain, i.e., TasNet, which achieved an impressive performance compared against the frequency-domain solutions. In \cite{8707067}, Yang \textit{et al.}  constructed a time-and-frequency (T-F) feature map by concatenating features for both time domain and frequency domain, and performed cross-domain joint embedding and clustering over this feature map, thus further improving the separation performance.

Despite considerable progress in recent years, the current literature has paid less attention to estimating the number of potential sources \cite{DBLP190403065,Higuchi2017,nachmani2020voice}, i.e., most of the above methods assume the number of speakers is known. The deep clustering approach requires the number of sources to cluster embeddings and obtain time-frequency (TF) masks\cite{Isik+2016}. The TasNet   needs the number of speakers to fix the dimension  of the output embedding, which makes it inflexible to deal with varying number  of sources \cite{8462116,8707065}. Actually, the number of speakers in realistic scenarios is often uncertain or even time-varying, thus accurately estimating the potential number of sources would be critical to subsequent separation. One way is to use the orthogonality of the high-dimensional embeddings. In \cite{Higuchi2017}, the number of speakers was estimated by computing the rank of the covariance matrix of high-dimensional embedding vectors. Although the existing works have made certain achievements in multi-channel scenarios \cite{ZHANG20171,9004553,gu2020multimodal}, there is much room for improving  source counting  and separation performance in single-channel scenarios. Consequently, sophisticated monaural speech separation models with unknown number of speakers are strongly required.

\begin{figure*}[t]
	\centering
	\includegraphics[width=0.99\textwidth]{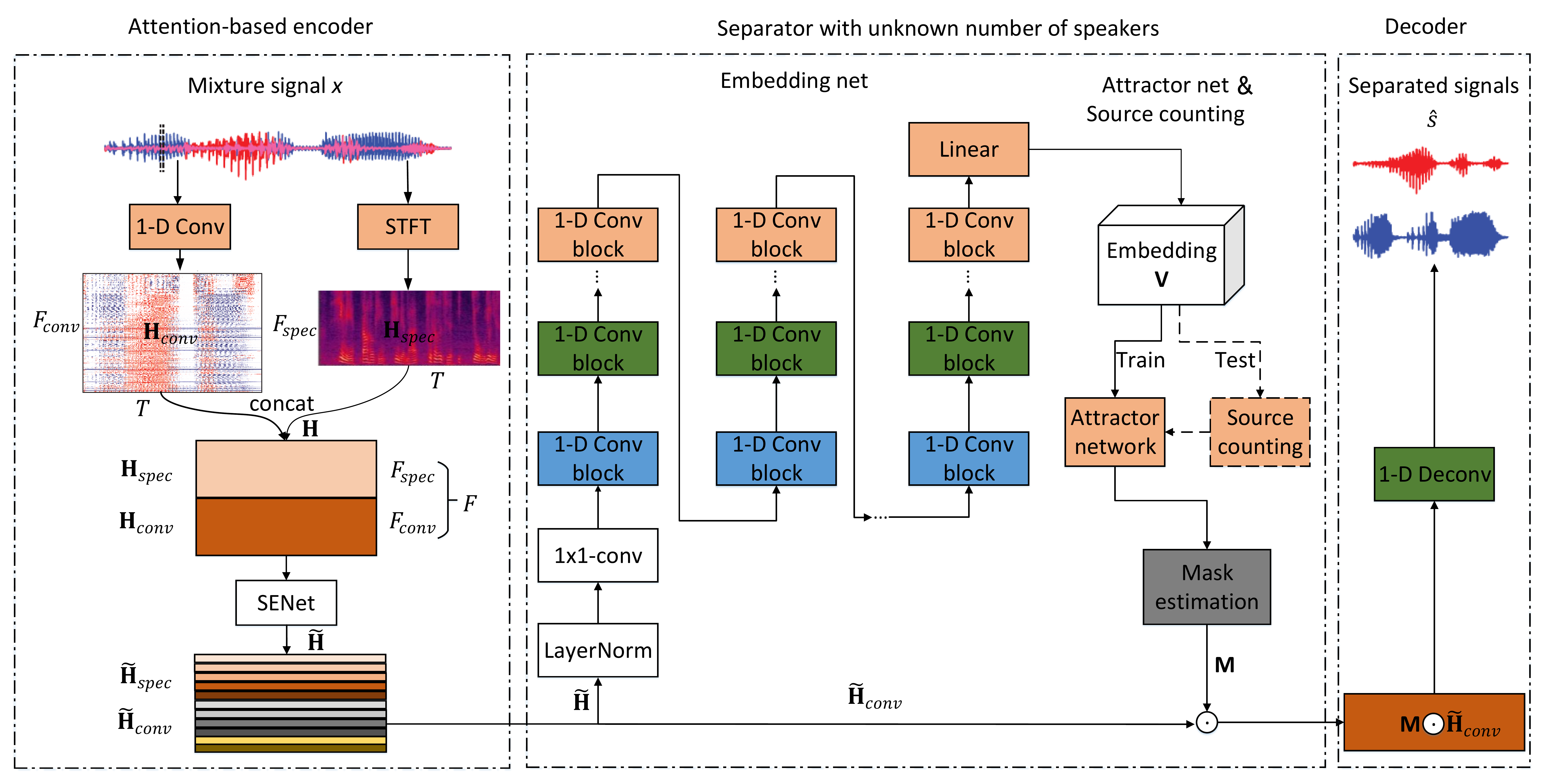}
	\vspace{-.5\baselineskip}
	\caption{ Architecture of the proposed method which consists of an attention-based encoder, a separator with unknown number of speakers, and a decoder.}
\end{figure*}

In this paper, we propose an attention-based T-F feature fusion  model for speech separation with unknown number of speakers. To further exploit the  T-F cross-domain feature, the attention mechanism  is adopted  in our encoding stage, which gives rise to better separation performance. Then, the encoded features are projected to high-dimensional vectors through an embedding part similar to TasNet. Theoretically, the orthogonal direction of above high-dimensional vectors is equivalent to the number of sources. Thus, the number of speakers could be estimated by computing the Gerschgorin disks of the embedding vectors. During the mask estimation stage, the ADANet \cite{8264702} is utilized to  estimate the mask of each source from the mixture using the similarity between the embeddings and each attractor. This implies that our network can be extended to an arbitrary number of sources once the attractors are established.  

\section{Proposed Method}

The flowchart of the proposed method is depicted in Fig. 1, which consists of three processing modules: an attention-based encoder, a separator with unknown number of speakers, and a decoder.  In the following subsections, each module of the proposed method is introduced in detail.


\subsection{ Attention-based Encoder  }

The monaural speech separation problem is formulated by estimating $C$ sources $ \{s_i(t)\}_{i=1,\cdots,C}$ given the monaural mixture $x(t) = \sum_{i=1} ^C s_i(t)$, where the number of speakers $C$ is unknown.
In the encoder part, we jointly utilize both the 1-dimensional (1-D) time-domain mixture and the 2-dimensional (2-D) frequency-domain spectrogram obtained by short-time Fourier transform (STFT). As shown in Fig. 1, the attention-based encoder first encodes the mixture $x(t)$ into a hybrid-domain 2-D feature map $\mathbf{H}$, which consists of $\mathbf{H}_{conv} $ and $\mathbf{H}_{spec} $  with $F = F_{conv} + F_{spec}$ frequency channels and $T$ time frames. The previous $F_{conv}$  channels are generated through the 1-D convolution operation  and  the subsequent $F_{spec}$ channels are obtained by the 2-D spectrogram. To integrate these two extracted features from T-F domains, the same window length and overlapping size for both domains are used.

In order to effectively exploit the T-F across-domain features $\mathbf{H}$, the squeeze-and-excitation network (SENet) \cite{PMID31034408} is employed to selectively emphasize informative features and meanwhile suppress useless ones. As shown in Fig. 2, we squeeze the global time information into a channel/frequency descriptor. This is
achieved by using global average pooling to generate
channel/frequency-wise statistics. A statistic $\mathbf{z} \in \mathbb{R}^{F}$ is generated by shrinking $\mathbf{H}$ along its time dimension such that the $f$-th element of $\mathbf{z}$ is calculated by
\begin{equation}
\mathbf{z}_{f}=\mathcal{F}_{s q}\left(\mathbf{H}_{f}\right)=\frac{1}{T} \sum_{t=1}^{T} \mathbf{H}_{f}(t), \quad f \in 1,2,...,F
\end{equation}

The information aggregated in the squeeze operation is followed by a gating mechanism which
consists of a bottleneck with two fully-connected layers around a non-linearity ReLU, i.e., a dimensionality-reduction layer with reduction ratio $r$,  the ReLU and then a dimensionality-increasing layer returning to the channel dimension of the transformation output
\begin{equation}
\mathbf{u}=\mathcal{F}_{e x}(\mathbf{z}, \mathbf{W})=\sigma  \big(g(\mathbf{z}, \mathbf{W})\big)=\sigma \big(\mathbf{W}_{2} \delta\left(\mathbf{W}_{1} \mathbf{z}\right) \big),
\end{equation}
where $\delta$ refers to the ReLU function  \cite{Sanchez2013}, $\sigma$  refers to the Sigmoid function,  $\mathbf{W}_{1} \in \mathbb{R}^{\frac{F}{r} \times F}$ and $\mathbf{W}_{2} \in \mathbb{R}^{F \times \frac{F}{r}} .$  The selective T-F fusion features $\widetilde{\mathbf{H}}$ is obtained by rescaling $\mathbf{H}$ with the activations $\mathbf{u}=[u_1~ u_2~ \cdots ~ u_F]$
\begin{equation}
\widetilde{\mathbf{H}}_{f}=\mathcal{F}_{\text {scale}}\left(\mathbf{H}_{f}, {u}_{f}\right)={u}_{f} \mathbf{H}_{f}, \quad f \in 1,2,... ,F
\end{equation}
which is adopted as the input of the separator module.
 
 
 \begin{figure}[t]
	\centering
	\includegraphics[width=0.499\textwidth]{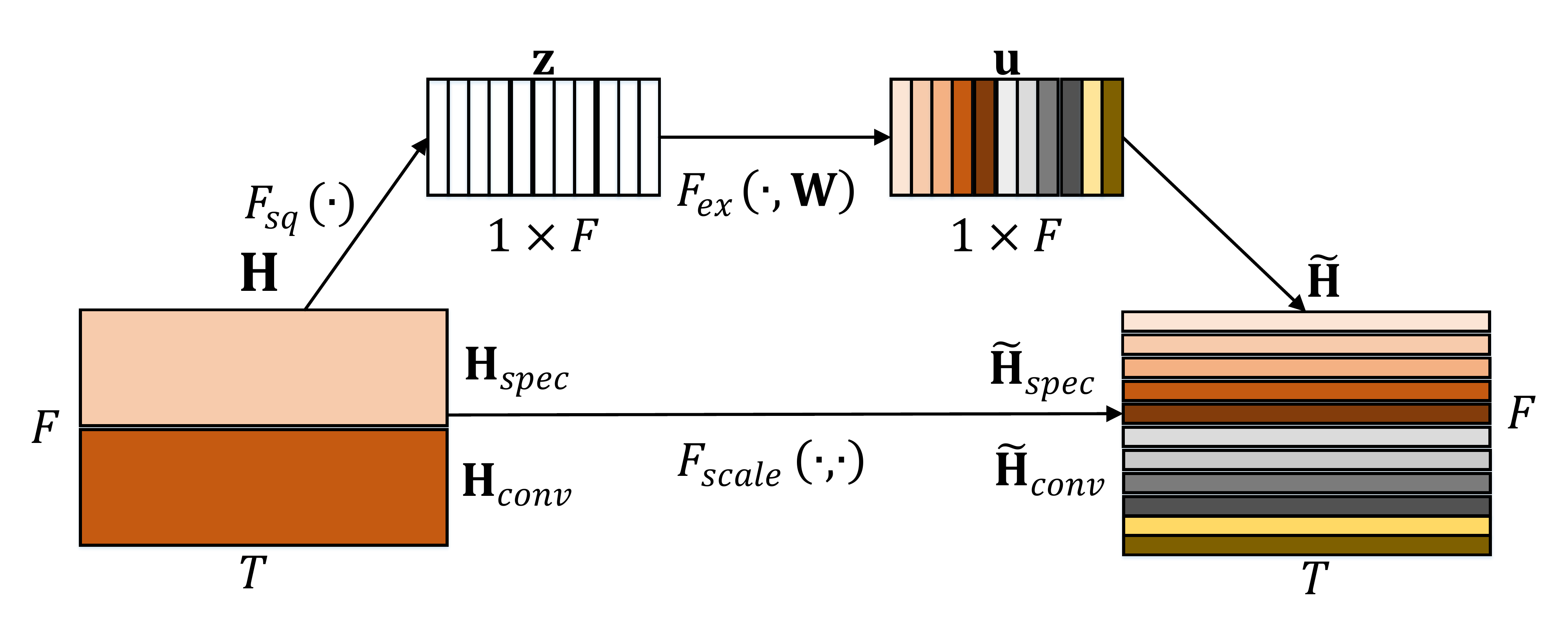}
	\vspace{-2.\baselineskip}
	\caption{ Process of emphasizing informative features using the SENet. }
\end{figure} 

\subsection{Separator with Unknown Number of Speakers }
The separator module contains three parts: an embedding network, an attractor network for mask estimation, and a source counting part. The overall proposed system allows us to separate the speech mixture without assuming the known number of speakers, which is automatically estimated by the source counting part.  

\subsubsection{Embedding Network}
In order to estimate the speaker assignment for each T-F index on
the hybrid-domain feature map $\widetilde{\mathbf{H}}$, we project the elements in  $\widetilde{\mathbf{H}}$ to $L$-dimensional embeddings $\mathbf{V} \in \mathbb{R}^{N \times L}$, where $N = T \times F_{conv}$, and $\mathbf{V}$ are in $C$ orthogonal directions through multiple layers of 1-D Conv blocks, as shown in Fig. 1, where the 1-D Conv block is actually a residual block consisting of a 1x1-conv, a dilated depth-wise convolution and a 1x1-conv module \cite{8707065}. Thus, the embeddings $\mathbf{V}$ based on $\widetilde{\mathbf{H}}$ are given as 
\begin{equation}
\mathbf{V} = Embed(\widetilde{\mathbf{H}}).
\end{equation}

\subsubsection{Deep Attractor Network for Mask Estimation}
To estimate masks of all the speakers in the mixture, we follow the ADANet in \cite{8264702} starting with $K$ initial centers $\{e_{k}\}_{k=1,\cdots,K}$.  By arbitrarily choosing $C$ out of the $K $ initial centers ($C$ is known in training but unknown in test time), we acquire $C$ new centroids by performing \textit{k}-means clustering with ${I}$ iterations on the embeddings $\mathbf{V}$. Considering there are all  $\tbinom{K}{C}$ possible selections out of the $K$ initial centers, we can obtain a total of  $\tbinom{K}{C}$ sets of centroids, among which we determine the set of centroids $\mathbf{A}$ with the largest in-set distance.  The masks for each speaker $\mathbf{M} \in \mathbb{R}^{T \times F_{conv} \times C}$ are then estimated by the dot product of the chosen centroids in $\mathbf{A}$ and the embeddings $\mathbf{V}$.

\subsubsection{Source Counting}
In the test stage, we estimate the number of sources based on   $\mathbf{V}$. Theoretically, the high-dimensional embedding vectors $\mathbf{V}$ after training are in $C$ directions that are orthogonal to each other. This property allows us to estimate the number of sources by estimating the rank of the covariance matrix of $\mathbf{V}$ \cite{Higuchi2017}. However, the above  strategy might be sensitive to noise intensity. To overcome this problem, we propose to utilize the Gerschgorin disk estimation (GDE) algorithm  \cite{Dong2013A}  to count the number of speakers. At first, we compute the covariance matrix of $\mathbf{V}=[\mathbf{v}_{1}~ \mathbf{v}_{2} ~\cdots ~\mathbf{v}_{N}]$
\begin{equation}
\begin{aligned}
\mathbf{B} \! =\!\frac{1}{N} \!\sum_{n=1}^N \!\mathbf{v}_{n} \!\mathbf{v}_{n}^{T} \! = \!\left(\begin{array}{cccc}
\!r_{11} & r_{12} & \cdots & r_{1 L}\! \\
\!r_{21} & r_{22} & \cdots & r_{2 L} \!\\
\!\vdots & \vdots & \ddots & \vdots \!\\
\!r_{L 1} & r_{L 2} & \cdots & r_{L L}\!
\end{array}\right) \! =\!\Bigg[\!\!\begin{array}{cc}
\!\mathbf{R}_{1}\! & \!\mathbf{r}\! \\
\!\mathbf{r}^{H} \!& \!r_{L L}\!
\end{array}\!\!\Bigg],
\end{aligned}\notag
\end{equation}
where $T$ and $H$ denote  transpose and conjugate transpose operators, respectively. $L$ is the dimension of $\mathbf{v}_n$, $\mathbf{R}_1$ is an $(L-1)\times (L-1)$ sub-matrix obtained by deleting the last row and column of $\mathbf{B}$, and $\mathbf{r}=\left[r_{1 L}, \ldots, r_{(L-1) L}\right]^{T}$. The eigenvectors $\mathbf{U}_{1}$ of $\mathbf{R}_1$ can be obtained via eigenvalue decomposition. 
Then we formulate a new matrix as 
\begin{equation}
\mathbf{U}_{2}=\left[\begin{array}{cc}
{\mathbf{U}_{1}} & {\mathbf{0}_{(L-1) \times 1}} \\
{\mathbf{0}_{1 \times(L-1)}} & {1}
\end{array}\right],
\end{equation}
based on which we transform $\mathbf{B}$ as below
\begin{equation}
\begin{aligned}
\mathbf{R}_{2} = \mathbf{U}_{2}^{H} \mathbf{B} \mathbf{U}_{2}=\left(\begin{array}{ccccc}
\lambda_{1} & 0 & \cdots & 0 & \rho_{1} \\
0 & \lambda_{2} & \cdots & 0 & \rho_{2} \\
\vdots & \vdots & \ddots & \vdots & \vdots \\
0 & 0 & \cdots & \lambda_{L-1} & {\rho}_{L-1} \\
{\rho}_{1}^{*} & {\rho}_{2}^{*} & \cdots & {\rho}_{L-1}^{*} & r_{L L}
\end{array}\right),
\end{aligned}
\end{equation}
where $*$ denotes conjugate operator, $\rho_{l}, \lambda_{l}$ ($l=1,2,...,L-1$) are radii and centers of the Gerschgorin disks, respectively. It is believed that the smaller  are  the  radii and centers of noise Gerschgorin disks, the larger are the radii and centers of remaining Gerschgorin disks which correspond to signal Gerschgorin disks. Therefore, we use Gerschgorin disk radii to estimate the number of sources through
\begin{equation}
G D E(k)=\left|\rho_{k}\right|-\frac{F_{\text{GDE}}(N)}{L-1} \sum_{l=1}^{L-1}\left|\rho_{l}\right|,  ~ k \in 1,2,...,L-1
\end{equation}
where $F_{\text{GDE}}(N)$ is an adjustable factor and also a non-increasing function of the sample size $N$. By detecting the first non-positive value $G D E\left(k_{0}\right)$, we can estimate the  number of speakers as $\hat{C}=k_{0}-1$, which is more robust to noise. 

\begin{table*}[t] 
\renewcommand{\arraystretch}{1.15}
	\centering\small
	\begin{threeparttable}
		\caption{ Separation performance comparison of different models trained on GRID  dataset in terms of SI-SNRi  and SDRi. 
		}
		\begin{tabular}{p{3.5cm}<{\centering}p{4.9cm}<{\centering}p{1.5cm}<{\centering}p{1.8cm}<{\centering}p{1.5cm}<{\centering}p{1.8cm}<{\centering}}
			\toprule 
			\multicolumn{2}{c}{\multirow{2}{*}{Methods}}   &   \multicolumn{2}{c}{ Two-Speaker Mixtures} & \multicolumn{2}{c}{ Three-Speaker Mixtures} \\
			&  & \makecell[c]{SDRi (dB)}&SI-SNRi (dB)  &    SDRi (dB)&SI-SNRi (dB) \\
			\midrule 
			\midrule
			\multirow{5}{*}{Two-Speaker Model}	&	DPCL++ \cite{Isik+2016} &   10.3&10.1 & 	\multicolumn{2}{c}{\multirow{5}{*}{Not Applicable}}	  \\  \cline{3-4}
			
			&	ADANet \cite{8264702}  &     10.1 &9.8     &    & \\ \cline{3-4}
			& Conv-TasNet-gLN \cite{8707065}&   14.4 &14.1  &    &\\ \cline{3-4}
			& Improved	\cite{8707067}&     15.4 &15.1   &      & \\ \cline{3-4}
			
			&$\mathbf{Proposed}$&   $\mathbf{16.2}$ & $\mathbf{16.0}$     &   &\\
			\midrule
			\midrule
			\multirow{5}{*}{Three-Speaker Model}	&		DPCL++ \cite{Isik+2016} &   	\multicolumn{2}{c}{\multirow{5}{*}{Not Applicable}} &  6.9 &  6.7 \\ \cline{5-6} 
			
			&ADANet \cite{8264702}&    & &   8.6 &      8.3 \\ \cline{5-6}
			&Conv-TasNet-gLN \cite{8707065}&   &  &      11.4 &    11.0\\ \cline{5-6}
			&Improved \cite{8707067}&   & &    12.3 &       12.0\\ \cline{5-6}
			
			&$\mathbf{Proposed}$&    & &    $\mathbf{13.4}$ &      $\mathbf{13.2}$ \\ 
			\midrule
			\midrule
			\multirow{3}{*}{Two \& Three-Speaker Model}	&	DPCL++ \cite{Isik+2016}&    10.7& 10.4 &    7.3 &    7.1 \\ \cline{3-6} 
			
			&	ADANet \cite{8264702}&  10.4 &10.2 &     8.5 &      8.2 \\ \cline{3-6}
			& $\mathbf{Proposed}$&       $\mathbf{15.5}$ & $\mathbf{15.3}$ &        $\mathbf{15.0}$ &      $\mathbf{14.5}$ \\ 
			
			\bottomrule 
		\end{tabular}
		
	\end{threeparttable}
\end{table*}

\subsection{Decoder}
After multiplying the feature map $\widetilde{\mathbf{H}}_{conv}$ by the estimated masks $\mathbf{M},$ we disassemble the masked encoded features into their original components. As shown in Fig. 1, the convolutional feature is dealt with through a deconvolution layer followed by overlap-add method to reconstruct the original signals $
\hat{s} =$ Decoder $ (\mathbf{M} \odot \widetilde{\mathbf{H}}_{conv})
$, where $\odot$ is element-wise multiplication. We choose the negative signal-to-distortion ratio as our training objective
\begin{equation}
\begin{aligned}
\mathcal{L}_{o s s} =
-10 \log _{10} \frac{\langle s, \hat{s}\rangle^{2}}{\|s\|^{2}\|\hat{s}\|^{2}-\left\langle s, \hat{s} \rangle^{2}\right.}
\end{aligned},
\end{equation}
where $\langle\cdot, \cdot\rangle$ is dot product,  and $\|s\|^{2}$ denotes the signal power.

\section{Experimental Evaluation}
We evaluate the proposed method on  2-speaker and 3-speaker mixtures, which are derived from the GRID dataset  \cite{12111}. The 30 hours of training set and 10 hours of validation set are  generated by different speakers from GRID  at various signal-to-noise ratio (SNR) from -2.5 dB to 2.5 dB. The 5 hours of test set is similarly generated with the validation set except that the speakers are different. All the speech waveforms are resampled to $8$ {kHz}. The window size of the STFT and the kernel size for the convolution layer in the encoder are both 2.5 ms, and the square root Hann window is used for STFT. 20-point DFT is performed to extract the 11-D log magnitude feature, combined with the 256-D feature extracted by the 1-D Conv, thus forming 267-D feature in $\mathbf{H}$. Then, the attention-based feature $\widetilde{\mathbf{H}}$ is gained by emphasizing different channels via SENet, where the reduction ratio $r$ is set to 16 \cite{PMID31034408}.   For the separator, the feature $\widetilde{\mathbf{H}}$  first goes through a $1 \times 1$-conv with 256 filters, followed by 8 residual 1-D Conv blocks, with dilated rate of $1,2, \ldots, 128,$ repeated for 4 times. $L=20$ is chosen as the embedding dimension  for better comparison \cite{7471631,Isik+2016,7952155,8264702,8462507}. We set $K=4$ initial centers and $I=1$ iteration for \textit{k}-means \cite{8264702}. The networks are trained  for 100 epochs using Adam algorithm with permutation invariant training \cite{7979557,7952154}. The training is performed end-to-end so that all components are jointly learned.

\subsection{Source Separation  Evaluation}

The proposed method is compared with other state-of-the-art methods \cite{Isik+2016,8264702,8707065,8707067} on the generated test set. These methods are categorized into three groups: the two-speaker model trained for two-speaker separation task, the three-speaker model trained for three-speaker separation task, and the two \& three-speaker model which is trained so that it can be applied to both two-speaker and three-speaker separation tasks with a unified model.

The separation results of different methods in terms of signal-to-distortion ratio improvement (SDRi)  and the scale-invariant SNR improvement (SI-SNRi) \cite{8264702,8707065,1643671} are shown in Table I. In the cases where the number of speakers in the mixture is different from that of the model target, we mark them as 'Not Applicable'. It is observed that the proposed method achieves the best performance on both SI-SNRi and SDRi through 2-speaker and 3-speaker model respectively. The proposed method solves the   source separation problem in time domain similar to Conv-TasNet-gLN \cite{8707065}, thus having much better separation performance compared with DPCL++ \cite{Isik+2016} and ADANet \cite{8264702} which formulate the separation problem in frequency domain.  However, different from Conv-TasNet-gLN \cite{8707065}, the proposed method  incorporates the ADANet for generating mask which is proved to be more effective for speech separation \cite{8707067}. More importantly, the proposed network with a single model enables us to handle the speech separation problem with different number of speakers.

Compared to the improved method \cite{8707067},  the attention module in our encoder part can emphasize more important information which contributes to better  separation performance. In addition, it is seen  from the results in 2 \& 3-speaker model that our proposed method  obtains a great improvement, i.e., more than 4 dB gains of SDRi and SI-SNRi on 2-speaker mixtures, and 6.5 dB gains of SDRi and SI-SNRi on 3-speaker mixtures compared with	DPCL++ \cite{Isik+2016} and ADANet \cite{8264702}. As a result, our model can obtain state-of-the-art separation results on both 2-speaker and 3-speaker mixtures in  a single model. It  should be also noted that our  2 \& 3-speaker model can obtain better performance on 3-speaker mixtures than that using our 3-speaker model. It suggests that  parameters of the proposed model learned in 2-speaker mixtures  contribute to the separation of 3-speaker mixtures, which can be used for studying speech separation with more speakers.

\subsection{Source Counting Evaluation}

The GDE algorithm is adopted based on the high-dimensional features to count the number of speakers. The  method of estimating the rank  of  the
covariance matrix $\mathbf{B}$ in \cite{Higuchi2017} is used as a comparison. For both methods, the  high-dimensional features $\mathbf{V}$ obtained from the mixture are used as input features. The test set consists of both 2-speaker and 3-speaker mixtures which are randomly selected from 3000 samples. 
We tune the threshold to  achieve the best source counting performance of the rank estimation method. As shown in Table II, the
proposed method identifies the number of speakers more accurately compared with the rank estimation method, and moreover it does not need to tune threshold for different number of speakers. The experimental results in Table I and Table II confirm the suitability for estimating the necessary information on the number of sources, which is often assumed to be known in advance.

\begin{table}[t] 
	\centering\small
	\caption{Source counting accuracy}
	\begin{tabular}
	{p{2.5cm}<{\centering}p{1.5cm}<{\centering}p{1.5cm}<{\centering}p{1.5cm}<{\centering}}
		\toprule\toprule 
		{\multirow{2}{*}{Methods}}   &   \multicolumn{3}{c}{Source Counting Accuracy $[\%]$} \\
		& Two Speakers & Three Speakers & Avg.  \\
		\midrule 
		{Rank Esti. \cite{Higuchi2017}} &   84.9 & 75.2 & 80.1 \\ 
		\midrule
		$\mathbf{Proposed}$	&  $\mathbf{95.7}$& $\mathbf{97.6}$&$\mathbf{96.7}$ \\ 
		\bottomrule 
	\end{tabular}
\end{table}

\section{Conclusion}
In this paper, we propose an attention-based network for source counting and separation in T-F fusion domain. The SENet is incorporated   into the encoder part of our model to emphasize more useful separation information, and  the source number is counted by computing the Gerschgorin disks based on the covariance matrix of embedding vectors. The ADANet  is then followed for mask estimation and enables our network to handle the speech separation with different number of speakers  in a single model. Experimental results show that our proposed method  can not only  separate a mixture of different numbers of speakers, but also can accurately detect the number of speakers. Most current single-channel speech separation methods only address the mixture data up to three speakers, and our method shows significant performance. However, for the separation of more speakers, the performance of our method might degrade. It has been concluded from Table I that the parameters of the proposed model learned in  fewer speaker mixtures would be beneficial to handle the mixture separation with more speakers, which motivates us to follow this direction to further investigate  multi-speaker speech separation.

\bibliographystyle{IEEEtran}
\bibliography{mybib}

\end{document}